\documentstyle[aps,preprint,epsf]{revtex}
\newcommand{\beq}{\begin{equation}}
\newcommand{\eeq}{\end{equation}}
\begin{document}
\draft
\tightenlines

\title{ Transition from non Fermi Liquid Behavior to Landau Fermi
Liquid Behavior Induced by Magnetic Fields}

\author{Yu.G. Pogorelov$^a$ and V.R. Shaginyan$^b$
\footnote{E-mail: vrshag@thd.pnpi.spb.ru}\\
$^a${\it Departamento de F\'{\i}sica, Universidade do Porto, 4150
Porto, Portugal}\\ $^b${\it Petersburg Nuclear Physics Institute,
RAS, Gatchina 188300, Russia }} \maketitle

\begin{abstract}
We show that a strongly correlated Fermi system with the fermion
condensate, which exhibits strong deviations from Landau Fermi
liquid behavior, is driven into the Landau Fermi liquid by applying
a small magnetic field $B$ at temperature $T=0$. This field-induced
Landau Fermi liquid behavior provides the constancy of the Kadowaki-Woods
ratio. A re-entrance into the strongly correlated regime is observed
if the magnetic field $B$ decreases to zero, then the effective mass
$M^*$ diverges as $M^*\propto 1/\sqrt{B}$. At finite temperatures,
the strongly correlated regime is restored at some temperature
$T^*\propto\sqrt{B}$. This behavior is of general form and takes place
in both three dimensional and two dimensional strongly correlated systems.
We demonstrate that the observed $1/\sqrt{B}$ divergence of the effective
mass and other specific features of heavy-fermion metals are accounted
for by our consideration.

\end{abstract}\bigskip

Recently, a peculiar critical point was observed in heavy-fermion
metal YbRh$_2$Si$_2$ at low temperatures $T$ \cite{gen}. This critical
point is driven by the magnetic field $B$ which suppresses the
antiferromagnetic order when reaches the critical value, $ B=B_{c0}$,
while the effective mass $ M^{*}$ diverges as $M^{*}\propto 1/
\sqrt{B-B_{c0}}$ \cite{gen}. The study of the magnetic field dependence
of the coefficients $A$, $\gamma_{0}$, and $\chi_{0}$ in the
resistivity, $\rho(T)=\rho_0+\Delta \rho$, with
$\Delta \rho =A(B)T^{2}$, specific heat, $C/T=\gamma_{0}(B)$, and the
magnetic $ac$ susceptibility, has revealed that YbRh$_2$Si$_2$
behaves as a true Landau Fermi liquid (LFL) for $B>B_{c0}$ and the
well-known Kadowaki-Woods ratio $A/\gamma_0$ \cite{kadw} is preserved
\cite{gen}. In contrast, for $B=0$, YbRh$_2$Si$_2$ demonstrates a
non Fermi liquid (NFL) behavior, and the resistivity follows a
quasilinear temperature dependence $\Delta\rho\sim T$ down to 80 mK
at which antiferromagnetic (AF) order takes place. At lower
temperatures, the resistivity in AF-ordered state is described by
$\Delta\rho\sim T^2$. A similar picture is observed in heavy-fermion
compounds Ce$M$In$_5$ ($M$=Ir, Co, and Rh), where the electronic
specific heat $C$ shows more pronounced metallic behavior at
sufficiently high magnetic fields \cite{kim}. These observations are
consistent with the de Haas van Alphen (dHvA) studies of CeIrIn$_5$,
which find that the effective mass decreases with increasing $B$
\cite{kim,haga}.

It is pertinent to note that heavy fermion metals are more likely three
dimensional (3D) then two dimensional (2D). The origin of NFL behavior
observed in heavy-fermion metals is still a subject of controversy
\cite{col}. Moreover, the observed constancy of the Kadowaki-Woods ratio
when $B\to B_{c0}$ \cite{gen} leads to the failure of the standard model
of heavy-fermion metals, when the mass renormalization is supposed to come
from the exchange by soft magnetic fluctuations in a 2D spin fluid
\cite{col1}. As a result, we are left with even more complicated and
challenging problems in the physics of strongly correlated electrons.

In this Letter, we study the nature of the critical behavior, assuming
that the fermion condensation phase transition (FCQPT) \cite{ms} plays
the role of the critical point. Analyzing the appearance of the
fermion condensate (FC) which occurs beyond the point of FCQPT in
an electron Fermi liquid and induces the NFL behavior, we show that
the liquid is driven, by applying a weak magnetic field, back into a
specific LFL state with effective mass $M^*\propto 1/\sqrt{B}$. The
LFL behavior induced by rather low magnetic fields, provides
constancy of the Kadowaki-Woods ratio. But the strongly correlated
regime is restored when the magnetic field $B$ decreases to zero,
while the effective mass $M^*$ diverges as $M^*\propto 1/\sqrt{B}$.
Also the strongly correlated regime is restored at some finite
temperature $T^*(B)\propto \sqrt{B}$. Such a behavior is of general
form and takes place in both three dimensional and two dimensional
strongly correlated systems. We demonstrate that the observed
crossover from NFL to LFL in certain heavy-fermion metals is
accounted for by our consideration.

For the reader's convenience we first outline the NFL behavior of
Fermi systems with FC and the main properties of FCQPT using, as
an example, a two-dimensional electron liquid in the superconducting
state induced by FCQPT \cite{ms,ams}. At $T=0$, the ground state
energy $E_{gs}[\kappa({\bf p}),n({\bf p})]$ is a functional of the
superconducting order parameter $\kappa({\bf p})$ and of the
quasiparticle occupation function  $n({\bf p})$ and is determined by
the known equation of the weak-coupling theory of superconductivity
(see e.g.  \cite{til})
\begin{equation}
E_{gs}=E[n({\bf p})] +\int \lambda_0V({\bf p}_1,{\bf p}_2)
\kappa({\bf p}_1) \kappa^*({\bf p}_2) \frac{d{\bf p}_1d{\bf p}_2}
{(2\pi)^4}.
\end{equation}
Here  $E[n({\bf p})]$ is the ground-state energy of normal Fermi
liquid, $n({\bf p})=v^2({\bf p})$ and $\kappa({\bf p})=v({\bf p})
\sqrt{1-v^2({\bf p})}$.  It is assumed that the pairing interaction
$\lambda_0V({\bf p}_1,{\bf p}_2)$ is weak. Minimizing $E_{gs}$ with
respect to $\kappa({\bf p})$ we obtain the equation connecting the
single-particle energy $\varepsilon({\bf p})$ to the superconducting
gap $\Delta({\bf p})$
\begin{equation}
\varepsilon({\bf p})-\mu=\Delta({\bf p})\frac{1-2v^2({\bf p})}
{2\kappa({\bf p})},
\end{equation}
including the chemical potential $\mu$. Here the single-particle energy
$\varepsilon({\bf p})$ is determined by the Landau equation \cite{lan}
\begin{equation}
\varepsilon({\bf p})=\frac{\delta
E[n({\bf p})]}{\delta n({\bf p})},
\end{equation}
while the equation for the superconducting gap $\Delta({\bf p})$ takes
the form
\begin{equation}
\Delta({\bf p})
=-\lambda_0\int V({\bf p},{\bf p}_1)\kappa({\bf p}_1)
\frac{d{\bf p}_1}{4\pi^2}.
\end{equation}
If $\lambda_0\to 0$, then, the maximum value of the superconducting
gap $\Delta_1\to 0$, and Eq. (2) reduces to that proposed in Ref. \cite{ks}
\begin{equation}
\varepsilon({\bf p})-\mu=0,\,\,\,{\rm if}\,\,\,\kappa ({ p})
\neq 0\,\,(0<n({ p})<1)\,\,\,{\rm for}\,\,\,{ p}\in
L_{{ FC}}:\,p_{i}\leq p\leq p_{f}.
\end{equation}
At $T=0$, Eq. (5) defines a new state of Fermi liquid with FC, such that
the modulus of the order parameter $|\kappa({\bf p})|$ has finite values
in the FC range of momenta $L_{ FC}$:  $p_i\leq p\leq p_f$, while
the superconducting gap can be infinitely small, $\Delta_1\to 0$, in
this range \cite{ms,ks,vol}. Such a state can be considered as
superconducting, with infinitely small value of $\Delta_1$ so that
the entropy of this state is zero. This state, created by the quantum phase
transition, disappears at $T>0$. The FCQPT can be considered as a
``pure'' quantum phase transition because it cannot take place at
finite temperatures. Generally, this quantum critical point should not
represent the termination at $T>0$ of a line of continuous transitions.
However it corresponds to a certain critical value of density $x=x_{ FC}$
(the critical point of FCQPT) which is determined also by Eq. (5).
Notice that at finite temperatures the FCQPT continues to have a strong
impact on the system properties, up to a certain temperature $T_f$ above
which FC effects become insignificant \cite{ms,dkss}. FCQPT does not
violate rotational or translational symmetry of the order parameter
$\kappa({\bf p})$. It follows from Eq. (5) that the quasiparticle
system ``splits'' into two quasiparticle subsystems: one in the
$L_{ FC}$ range, occupied by the quasiparticles with enhanced
effective mass $M^*_{FC}\propto 1/\Delta_1$, and another with LFL
effective mass $M^*_{L}$ at $p<p_i$. If $\lambda_0\neq0$, then
$\Delta_1$ becomes finite, and the finite value of the effective mass
$M^*_{FC}$ in $L_{FC}$ can be obtained from Eq. (2) as
\cite{ms,ams} \begin{equation} M^*_{FC} \simeq p_{
F}\frac{p_f-p_i}{2\Delta_1}, \end{equation} while the effective mass
$M^*_{L}$ is only weakly disturbed.  Here $p_{F}$ is the
Fermi momentum.  It follows from Eq. (6) that the quasiparticle
dispersion can be presented by two straight lines characterized by
the effective masses $M^*_{FC}$ and $M^*_{L}$ respectively.
These lines intersect near the electron binding energy $E_0$ which
defines an intrinsic energy scale of the system:  \begin{equation}
E_0=\varepsilon({\bf p}_f)-\varepsilon({\bf p}_i)
\simeq\frac{(p_f-p_i)p_F}{M^*_{FC}}\simeq 2\Delta_1.
\end{equation}
Let us assume that FC has just taken place, that is $p_i \to p_{F}
\leftarrow p_f$, the deviation $\delta n(p)$ from LFL occupation
function is small (though finite), and $\lambda_0 \to 0$. Expanding the
functional $E[n(p)]$ in Eq. (3) in Taylor's series with respect to
$\delta n(p)$ and retaining the leading terms, we have
\begin{equation}
\Delta E=\sum_{\sigma_1}\int\varepsilon_0({\bf p_1},\sigma_1)
\delta n({\bf p_1},\sigma_1)\frac{d{\bf p}_1}{(2\pi)^2}+
\sum_{\sigma_1\sigma_2}\int F_L({\bf p}_1,
{\bf p}_2,\sigma_1,\sigma_2)
\delta n({\bf p_1},\sigma_1)
\delta n({\bf p_2},\sigma_2)
\frac{d{\bf p}_1 d{\bf p}_2}{(2\pi)^4},
\end{equation}
where $F_L({\bf p}_1,{\bf p}_2,\sigma_1,\sigma_2)=
\delta^2 E/\delta n({\bf
p}_1,\sigma_1)\delta n({\bf p}_2,\sigma_2)$ is the Landau interaction,
and $\sigma$ denotes the spin states. Varying both sides of Eq. (8)
with respect to the functions $\delta n(p)$ and taking into account
Eq. (5), one obtains the FC equation
\begin{equation}
\mu=\varepsilon({\bf p},\sigma) =
\varepsilon_0({\bf p},\sigma)+\sum_{\sigma_1}\int F_L({\bf p},{\bf
p}_1,\sigma,\sigma_1)\delta n({\bf p_1},\sigma_1) \frac{d{\bf
p}_1}{(2\pi)^2};\,\,\: p_i\leq p\leq p_f\in L_{FC}.
\end{equation}
Equation (9) acquires non-trivial solutions at the density $x=x_{FC}$
if the Landau amplitude $F_{ L}$ (depending on density) is positive
and sufficiently large, so that the potential energy integral on the
right hand side of Eq. (9) prevails over the kinetic energy
$\varepsilon_0({\bf p})$ \cite{ks}.  Note, that in case of heavy fermion
metals, this condition can be easily satisfied because of the huge
effective mass. It is also seen from Eq. (9) that the FC
quasiparticles form a collective state, since their energies are
defined by the macroscopical number of quasiparticles within $L_{ FC}$,
and vice versa. The shape of their spectrum is not affected by the
Landau interaction, which, generally speaking, depends on the system
properties, including the collective states, impurities, etc. The only
thing defined by the interaction is the width $p_i-p_f$ of $L_{ FC}$
(provided it exists). Thus, we can conclude that the spectra related
to FC are of universal form.

At temperatures $T\geq T_c$ when $\Delta _{1}$ disappears, Eq. (6) for
the effective mass $M^*_{ FC}$ is changed for \cite{ms,ams}
\begin{equation}
M^*_{FC}\simeq p_F\frac{p_f-p_i}{4T}.
\end{equation}
The energy scale separating the slow dispersing low energy part,
defined by the effective mass $M^*_{ FC}$, from the faster
dispersing relatively high energy part, defined by the effective
mass $M^*_{ L}$, can be estimated as $E_{0}\simeq p_{{ F}}
(p_f-p_i)/M^{*}_{{ FC}}$ \cite{ms,ams}, so for the case of Eq.
(10) it is
\begin{equation}
E_0\simeq 4T.
\end{equation}
It follows from Eq. (10) that $M^*_{FC}$ depends on the temperature,
and the width $\gamma$ of the single-particle excitations results
$\gamma\sim T$, leading to a linear temperature dependence
$\Delta\rho\sim T$ \cite{dkss}. This contrasts with the well-know LFL
relations:  $\gamma\sim T^2$, and $\Delta\rho\sim T^2$.

Now we consider the behavior of an electronic system with FC in magnetic
fields, supposing the coupling constant $\lambda_0\neq 0$ to be infinitely
small. As we have seen above, at $T=0$, the superconducting order
parameter $\kappa({\bf p})$ is finite in the FC range, while the maximum
value of the superconducting gap $\Delta_1\propto \lambda_0$ is
infinitely small. Therefore, any small magnetic field $B \neq 0$ will
destroy the coherence of $\kappa({\bf p})$ and thus FC itself. Also,
the existence of FC can not be compatible with the evident Zeeman
splitting of quasiparticle energy bands $\varepsilon({\bf p},\sigma)
= \varepsilon({\bf p})-\sigma \mu_{eff}B$ (see below for
$\mu_{eff}$).  To define the type of FC restructuring, simple energy
reasons are invoked. On the one hand, the energy gain $\Delta E_B$
due to the magnetic field $B$ is $\Delta E_B\propto B^2$ and tends to
zero with $B\to 0$. On the other hand, occupying the finite range
$L_{ FC}$ in the momentum space, FC delivers a finite gain in the
ground state energy \cite{ks}. Thus, a new state replacing FC should
be very close in its ground state energy to the former state. Such a
state is given by the multiconnected Fermi sphere, where the smooth
quasiparticle distribution function $n({\bf p})$ in the $L_{FC}$ range
is replaced by a multiconnected distribution $\nu({\bf p})$ \cite{asp}
\begin{equation}
\nu({\bf p})=\sum\limits_{k=1}^n\theta (p-p_{2k-1})\theta (p_{2k}-p),
\end{equation}
where the parameters $p_i\leq p_1<p_2<\ldots <p_{2n}\leq p_f$ are
adjusted to obey the normalization conditions:
\begin{equation}
\int_{p_{2k}}^{p_{2k+3}}\nu({\bf p})
\frac{d{\bf p}}{(2\pi)^3}=\int_{p_{2k}}^{p_{2k+3}}
n({\bf p})\frac{d{\bf p}}{(2\pi)^3}.
\end{equation}
For the definiteness sake, we consider the most interesting case of 3D
system, while the consideration of a 2D system goes along the same line.
We note that the idea of multiconnected Fermi sphere, with production
of new, interior segments of the Fermi surface, has been considered
already \cite{llvp,zb}. Let us assume that the thickness of each interior
block is approximately the same $p_{2k+1}-p_{2k}\simeq \delta p$ and
$\delta p$ is defined by $B$. Then, the single-particle energy in the
region $L_{FC}$ can be fitted by
\begin{equation}
\varepsilon({\bf p}) - \mu \sim \mu\frac{\delta p}{p_F}
\left [ \sin\left(\frac{p}{\delta p}\right) + b(p) \right ].
\end{equation}
The blocks are formed since all the single particle states around the
minimum values of the fast sine function are occupied and those
around its maximum values are empty, the average occupation being
controlled by a slow function
$b({\bf p})\approx\cos [\pi n({\bf p})]$. It is seen from Eq. (14)
that the effective mass $m^*$ at each
internal Fermi surface is of the order of the bare mass $m_0$,
$m^*\sim m_0$.  Upon substituting $n({\bf p})$ in Eq. (8) by
$\nu({\bf p})$, defined by Eqs. (12) and (13), and taking into
account the Simpson's rule, we obtain that the minimum loss in the
ground state energy due to the formation of blocks is about $(\delta
p)^4$.  This result can be understood if one bears in mind that the
continuous FC function $n({\bf p})$ delivers the minimum value to the
energy functional, Eq.  (8), while the approximation $\nu({\bf p})$
by steps of size $\delta p$ produces the minimum error of the order
of $(\delta p)^4$. On the other hand, this loss must be compensated
by the energy gain due to the magnetic field. Thus, we arrive at
\begin{equation} \delta p\propto \sqrt{B}.  \end{equation} With an
account taken of the Zeeman splitting in the dispersion law, Eq.
(14), each of the blocks is polarized, since its outer areas are
occupied only by spin-up quasiparticles. The width of this areas in
the momentum space $\delta p_0$ is given by \begin{equation}
\frac{p_F\delta p_0}{m^*}\sim B\mu_{eff},
\end{equation}
where $\mu_{eff}\sim \mu_B$ is the effective moment. We can consider
such a polarization, without perturbing the previous estimates, since
it is seen from Eq. (15) that $\delta p_0/\delta p\ll 1$. The total
polarization $\Delta P$ is obtained multiplying $\delta p_0$ by the
number $N$ of the blocks which is proportional to $1/\delta p$,
$N\sim (p_f-p_i)/\delta p$. Taking into account Eq. (15), we obtain
\begin{equation}
\Delta P\sim m^*\frac{p_f-p_i}{\delta p}B\mu_{eff}\propto \sqrt{B},
\end{equation}
and thus it prevails over $\sim B$ contribution from the LFL part.
On the other hand, this quantity can be expressed as
\begin{equation}
\Delta P\propto M^*B,
\end{equation}
where $M^*$ is the ``average'' effective mass related to the finite
density of states at the Fermi level,
\beq
M^*\sim N m^*\propto\frac{1}{\delta p}.
\eeq
We can also conclude that $M^*$ defines the specific heat.

Otherwise Eq. (15) can be examined, starting from a different point
surmising that multiconnected Fermi sphere can be approximated
by a single block. Let us put $\lambda_0=0$.
Then, the energy gain due to the
magnetic field $\Delta E_B\sim B^2M^*$. The energy loss
$\Delta E_{ FC}$ because of the restructuring of the FC state can
be estimated by using the Landau formula which directly follows from
Eqs. (8) and (9)
\begin{equation}
\Delta E_{ FC}=\int(\varepsilon({\bf p})-\mu)\delta
n({\bf p})\frac{d{\bf p}^3}{(2\pi)^3}.
\end{equation}
As we have seen above, the region occupied by variation $\delta
n({\bf p})$ has the length $\delta p$, while $(\varepsilon({\bf p})
-\mu)\sim (p-p_F)p_F/M^*$. As a result, we have, $\Delta E_{ FC}=
\delta p^2/M^*$. Upon equating $\Delta E_B$ and $\Delta E_{FC}$
and taking into account Eq. (19), we arrive at the following equation
\begin{equation}
\frac{\delta p^2}{M^*}\propto \delta p^3\propto
\frac{B^2}{\delta p},
\end{equation}
which coincides with Eq. (15).

It follows from Eqs. (17) and (18) that the effective mass
$M^*$ diverges as
\begin{equation}
M^*\propto \frac{1}{\sqrt{B}}.
\end{equation}
Equation (22) shows that, by applying a magnetic field $B$,
the system can be driven back into LFL with the effective
mass $M^*(B)$ dependent on the magnetic field. It was demonstrated
that the constancy of the Kadowaki-Woods ratio is obeyed by systems
in the strongly correlated regime when the effective mass is
sufficiently large \cite{ksch}. Therefore, we are led to the
conclusion that, by applying magnetic fields, the system is driven
back into LFL where the constancy of the Kadowaki-Woods ratio is
obeyed. Since the resistivity $\Delta\rho\propto (M^*)^2$
\cite{ksch}, we obtain from Eq. (22) \beq \Delta\rho\propto
\frac{1}{B}.\eeq At finite temperatures, the system persists to be
LFL, but there is a temperature $T^*(B)$ at which the polarized state
is destroyed. To calculate the function $T^*(B)$ , we observe that
the effective mass $M^*$ characterizing the single particle spectrum
cannot be changed at $T^*(B)$. In other words, at the crossover
point, we have to compare the effective mass defined by $T^*$, Eq.
(10), and that defined by the magnetic field $B$ \begin{equation}
\frac{1}{M^*}\propto T^*\propto \sqrt{B}.
\end{equation}
As a result, we obtain
\begin{equation}
T^*(B)\propto \sqrt{B}.
\end{equation}
At temperatures $T\geq T^*$, the system comes back into the state
with $M^*$ defined by Eq. (10), and we observe the NFL behavior.
It follows from Eq. (25), that a heavy fermion system at some
temperature $T$ can be driven back into LFL by applying strong enough
magnetic field $B\geq B_{cr}\propto (T^*)^2$.  We can also conclude,
that at finite temperature $T$, the effective mass of a heavy fermion
system is relatively field independent at magnetic fields $B\leq
B_{cr}$ and show a more pronounced metallic behavior at $B\geq
B_{cr}$ since the effective mass is decreased, see Eq. (22). The same
behavior of the effective mass can be observed in the dHvA
measurements. We note that our consideration is valid up to
temperatures $T\ll T_f$.

Now we are in position to consider the nature of the field-induced
quantum critical point in YbRh$_2$Si$_2$. The properties of this
antiferromagnetic (AF) heavy fermion metal with the ordering Ne\`{e}l
temperature $T_N=70$ mK were recently investigated in
Refs.\cite{gen,ishida}. In AF state, this metal
shows LFL behavior. As soon as the
weak AF order is suppressed either by a tiny volume expansion or by
temperature, pronounced deviations from LFL behavior are observed.
The experimental facts show that the spin density wave picture is
failed when dealing with the obtained data \cite{gen,col1,ishida}. We
assume that the electron density in YbRh$_2$Si$_2$ is close to
the critical value $(x_{FC}-x)/x_{FC}\ll1$ \cite{shag}, so that this
system can be easily driven across FCQPT. Then, in the AF state, the
effective mass is given by Eq. (22) and the electron system of
YbRh$_2$Si$_2$ possesses LFL behavior. When the AF state is
suppressed at $T>T_N$ the system comes back into NFL. By tuning
$T_N\to 0$ at a critical field $B=B_{c0}$, the itinerant AF order is
suppressed and replaced by spin fluctuations \cite{ishida}. Thus, we
can expect absence of any long-ranged magnetic order in this state,
and the situation corresponds to a paramagnetic system with strong
correlation in the field $B=0$.  As a result, the FC state is
restored and we can observe NFL behavior at any temperatures in
accordance with experimental facts \cite{gen}.  As soon as an excess
magnetic field $B>B_{c0}$ is applied, the system is driven back into
LFL. To describe the behavior of the effective mass, we can use Eq.
(22) substituting $B$ by $B-B_{c0}$
\begin{equation}
M^*\propto \frac{1}{\sqrt{B-B_{c0}}}.
\end{equation}
Equation (26) demonstrates the $1/\sqrt{B-B_{c0}}$ divergence of the
effective mass, and therefore the coefficients $\gamma_0(B)$ and
$\chi_{0}(B)$ should have the same behavior. Meanwhile the coefficient
$A(B)$ diverges as $1/(B-B_{c0})$, being proportional to $(M^*)^2$
\cite{ksch}, and thus preserving the Kadowaki-Woods ratio, in
agreement with experimental findings \cite{gen}. To construct $T-B$
phase diagram for YbRh$_2$Si$_2$ we use the same replacement $B\to
B-B_{c0}$ in Eq. (25) \begin{equation} T^*(B)\propto \sqrt{B-B_{c0}}.
\end{equation}
The phase diagram given by Eq. (27) is in good qualitative agreement
with the experimental data \cite{gen}. We recall that our
consideration is valid at temperatures $T \ll T_f$. The experimental
phase diagram shows that the behavior $T^*\propto \sqrt{B-B_{c0}}$ is
observed up to $150$ mK \cite{gen} and allows us to estimate the
magnitude of $T_f$ which can reach at least $1$ K in this system.
It pertinent to note, that it follows directly from our
consideration that the similar $T-B$ phase diagram given by Eq. (27)
can be observed at least in case of strongly overdoped
high-temperature compounds. Except very close to the small values of
both $B$ and $T$, because at $T\leq T_c$ the magnetic field is to be
$B>B_c$, here $B_c$ is the critical field suppressing the
superconductivity.

To conclude, we have demonstrated that a new type of the quantum
critical point observed in heavy-fermion metal YbRh$_2$Si$_2$ can
be identified as FCQPT.

This work was supported in part by the Russian Foundation for Basic
Research, No 01-02-17189.

\end{document}